\documentclass{ws-procs975x65}
\usepackage{amsmath}
\usepackage{amssymb}
\usepackage{amsfonts}
\usepackage{mathrsfs}
\usepackage[T1]{fontenc}
\usepackage[latin9]{inputenc}
\usepackage{units}
\usepackage[numbers]{natbib}

\begin{document}

\title{Polymerization, the Problem of Access to the Saddle Point Approximation, and Thermodynamics}
\author{Hugo A. Morales-T\'{e}cotl$^*$, Daniel H. Orozco-Borunda and Saeed Rastgoo$^\dagger$}

\address{Departamento de F\'{i}sica, Universidad Aut\'{o}noma Metropolitana - Iztapalapa\\ San Rafael Atlixco 186, Mexico D.F. 09340, Mexico\\ ${}^*$E-mail: hugo@xanum.uam.mx\\   
$^\dagger$E-mail: \email{saeed@xanum.uam.mx}}

\begin{abstract}
The saddle point approximation to the partition functions is an important
way of deriving the thermodynamical properties of black holes. However,
there are certain black hole models and some mathematically analog mechanical
models for which this method can not be applied directly. This is due to the
fact that their action evaluated on a classical solution is not finite
and its first variation does not vanish for all consistent boundary conditions. These problems can be dealt with
by adding a counter-term to the classical action, which is a solution
of the corresponding Hamilton-Jacobi equation. 

In this work however, we seek an alternative solution to this problem via the polymer quantization which is  motivated by the loop quantum gravity.

\end{abstract}


\bodymatter


\section{Introduction}

The thermodynamic of black holes can be studied through the saddle point approximation to the  Euclidean path integral of the system, since the latter  is interpreted as the partition function of the black hole in
canonical ensemble. However, there are a class of 2D dilatonic models of black holes for which this approximation method can not be applied (see Refs.~\refcite{Regge1974,Gibbons1977,H.Liebl1997,Papadimitriou2005,McNees2005,Mann2006}) due to the fact that the action, $S$, is not functionally differentiable or it is  not finite on classical solutions for all the variations
of the fields, compatible with the boundary and fall-off conditions \cite{Grumiller2007}. 

There are also analog mechanical models exhibiting the same technical problems that can be used as toy models to study this issue in a more simpler way \cite{Grumiller2007a}. The issue in both class of systems can be resolved by adding a boundary term $G$ to the action, turning it into an ``improved action'' \(\Gamma=S+G\), where $G$  is the Hamilton's principal function of the system.

As an alternative method of  dealing with this issue, we would like to apply the polymer quantization (see Ref.~\refcite{Ashtekar:2002sn}) to this problem.   This method of quantization uses techniques from loop quantum gravity \cite{Rovelli:2004tv,Thiemann:2007zz,GambiniPullin2011,Bojowald2011,RovelliVidotto2014}. The polymer quantization has been applied to a range of models yielding interesting and in many cases different results from the standard Schr\"odinger quantization
 \cite{Corichi:2006qf,Corichi:2007tf,Flores-Gonzalez:2013zuk,Hossain:2010eb,Garcia-Chung:2014sza, Cumsille:2015xaa,HusainWinkler2003,Ashtekar:2009dn,Ashtekar:2010gz,Ashtekar:2010ve,kajuri2014path, Parra:2014mxa,MoralesRuelas2015}.

To see the issue in a bit more detail, consider the semiclassical approximation for a system with the Euclidean path integral
\begin{equation}
\mathcal{Z}=\int\prod_{j}\mathscr{D}\phi_{j}\exp\left(-\frac{1}{\hbar}S_{E}[\phi_{j}]\right),\label{eq:path-int-gen}
\end{equation}
in which $\phi_{j}$ are the fields of the theory and $S_{E}$ is
its Euclidean action. Assuming that it is possible to expand the action around the classical
solutions $\phi_{j}\big|_{cl}$ as
\begin{equation}
S_{E}\left[\left.\phi_{j}\right|_{cl}+\delta\phi_{j}\right]=S_{E}\left[\left.\phi_{j}\right|_{cl}\right]+\delta S_{E}\left[\left.\phi_{j}\right|_{cl}+\delta\phi_{j}\right]+\frac{1}{2}\delta^{2}S_{E}\left[\left.\phi_{j}\right|_{cl}+\delta\phi_{j}\right]+\cdots,\label{eq:expnd-SE}
\end{equation}
one can then substitute this into (\ref{eq:path-int-gen}) and get 
\begin{equation}
\mathcal{Z}\approx\exp\left(-\frac{1}{\hbar}S_{E}\left[\left.\phi_{j}\right|_{cl}\right]\right)\int\prod_{j}\mathscr{D}\phi_{j}\exp\left(-\frac{1}{2\hbar}\delta^{2}S_{E}\left[\left.\phi_{j}\right|_{cl}+\delta\phi_{j}\right]\right),\label{eq:saddle-gen}
\end{equation}
up to the quadratic term. This is the saddle point approximation to the model which gives us
access to the semiclassical information about the system.

More precisely, a saddle point approximation (\ref{eq:saddle-gen}) of (\ref{eq:path-int-gen})
is possible only if the following conditions are met:
\begin{enumerate}
\item well-definedness of variational
principle: $S_{E}$ must be functionally differentiable for all the variations
of the fields, compatible with the boundary and fall-off conditions of the fields, so that any boundary term must vanish by virtue of these conditions and thus we can write $\delta S_{E}=\int d^{n}x\frac{\delta S_{E}}{\delta\phi_{j}}\delta\phi_{j}$. \item Given condition 1, then $S_{E}$ on classical solutions must remain finite.
\item Given conditions 1 and 2, the Gaussian integral 
in (\ref{eq:saddle-gen}) must remain finite. Also $\delta^{2}S_{E}\left[\left.\phi_{j}\right|_{cl}+\delta\phi_{j}\right]>0$ due to presence of a minus sign in the exponent in (\ref{eq:path-int-gen}).
\end{enumerate} 
\section{Analog Mechanical Model And Its Improved Action\label{sec:analog-model}}

There are several types of simpler analog models that exhibit the aforementioned ill-defined semiclassical approximation. One such class of models corresponds to a single particle systems
in  half binding potentials \cite{Grumiller2007a}. We choose
a simple system in this class with an inverse square potential with
the Euclidean action $S_E=\int d\tau\left(\frac{m}{2}\left(\frac{dq}{d\tau}\right)^{2}+\frac{k}{q^{2}}\right)$ with $\tau$ 
being the Euclidean time. Since at the boundary $\tau\rightarrow\infty$, we have $q_{cl}\rightarrow\infty$, this results in $S_E$ not being functionally differentiable or even if so,  $S_{E}\big|_{q_{cl}}\rightarrow\infty$ \cite{Grumiller2007a}. 

As mentioned before, one known and common way to solve this problem is to add a boundary term $-G$ to the action to get an improved action, $\Gamma=S_{E}-\left.G(q,t)\right|_{0}^{t_{f}}$ with $G$ being the Hamilton's principal function of the system. Then it can be shown that $\Gamma$  is functionally differentiable and $\Gamma\big|_{q_{cl}}<\infty$ \cite{Grumiller2007a}.

\section{Polymer Quantization Of The Model\label{sec:eff-poly-with-HJ}}

Polymer representation is a singular representation of the Weyl group  that is not unitarily equivalent to the Schr\"{o}dinger representation. This is because it does not obey the weak continuity condition in Stone-von Neumann theorem \cite{Corichi:2007tf}. In our case, one can consider two polarizations of this polymer representation
in which either \(q\) or \(p\) are discrete and the other variable in bounded. In the former case the basic operators are $\hat{q}$ and $\hat{V}_{\lambda}$ which are represented on the Hilbert space, $\mathcal{H}_{q_0}$, as 
\begin{equation}
\hat{q}|q_{n}\rangle=  q_{n}|q_{n}\rangle,\,\,\,\,\,\,\,\,\,\,\,\,\,\,\,\,\, \hat{V}_{\lambda}|q_{n}\rangle=  |q_{n}-\lambda\rangle,\label{eq:q-descrt-polarz-1}
\end{equation} 
$p$ can not be represented on $\mathcal{H}_{q_0}$ due to lack of weak continuity and the $q$ space is discrete with lattice points being \(q_n=q_0+n\lambda\) with $n\in\mathbb{Z}$ for some initial value $q_0$. In the latter case, the basic operators are 
\begin{equation}
\hat{U}_{\mu}|p_{n}\rangle=  |p_{n}-\mu\rangle,\,\,\,\,\,\,\,\,\,\,\,\,\,\,\,\,\, \hat{p}|p_{n}\rangle=  p_{n}|p_{n}\rangle,\label{eq:p-descrt-polarz-1}
\end{equation}
$q$ can not be represented on $\mathcal{H}_{p_0}$ and the $p$ space is discrete with lattice points being \(p_n=p_0+n\mu\) and $n\in\mathbb{Z}$. \subsection{The model with discrete $q$}

In this case 
the kinetic term of Euclidean Hamiltonian $H_{E}=\frac{p^{2}}{2m}-\frac{k}{q^2}$
can be represented as $p^{2}\rightarrow\widehat{p_{\lambda}^{2}}=\frac{\hbar^{2}}{\lambda^{2}}\left(2-\hat{V}_{\lambda}-\hat{V}_{-\lambda}\right)$ (see e.g. Ref.~\refcite{Flores-Gonzalez:2013zuk}). The potential can be represented using
a regularization following Thiemann (Ref.~\refcite{Thiemann1998}),
\begin{equation}
\frac{1}{\sqrt{|q|}}= \frac{2}{i\lambda}V_{-\lambda}\left\{ \sqrt{|q|},V_{\lambda}\right\}
=  \frac{V_{-\lambda}}{i\lambda}\left\{ \sqrt{|q|},V_{\lambda}\right\} +\left\{ \sqrt{|q|},V_{\lambda}\right\} \frac{V_{-\lambda}}{i\lambda},
\end{equation}
where we have chosen a specific symmetrization. It is obvious that other types of orderings are
also possible. The full quantum Euclidean Hamiltonian is then
\begin{equation}
\hat{H}_{E}=\frac{\hbar^{2}}{2m\lambda^{2}}\left(2-\hat{V}_{\lambda}-\hat{V}_{-\lambda}\right)-k\left(\frac{\hat{V}_{-\lambda}}{\lambda}\hbar\left[\widehat{\sqrt{|q|}},\hat{V}_{\lambda}\right]+\left[\widehat{\sqrt{|q|}},\hat{V}_{\lambda}\right]\hbar\frac{\hat{V}_{-\lambda}}{\lambda}\right)^{4}.\label{eq:Euclid-H-Qreps}
\end{equation}Using path integral methods one can derive the effective Hamiltonian (Ref.~\refcite{Morales-Poly2015})
\begin{equation}
H_{\textrm{eff}}=\frac{2\hbar^{2}}{m\lambda^{2}}\sin^{2}\left(\frac{\lambda p}{2\hbar}\right)-W^{T}=\frac{2\hbar^{2}}{m\lambda^{2}}\sin^{2}\left(\frac{\lambda p}{2\hbar}\right)-\frac{k\hbar^{4}}{\lambda^{4}}\left(\sqrt{|q-\lambda|}-\sqrt{|q+\lambda|}\right)^{4}.\label{eq:H-eff-Thiem}
\end{equation}
 It can be seen that the effective potential allows for $q_{cl}\big|_{\tau\rightarrow\infty}\rightarrow\infty$ and thus the boundary term in the action does not vanish and the issues with the saddle point approximation persist in this polarization. However it can be shown that the counter-term is modified as $C_{\textrm{poly}}=C^{(0)}+\sum_{n=1}\frac{\lambda^{2n}}{\hbar^{2n}}C^{(2n)}$.
The purely classical term here with $\lambda=0$ matches the classical Hamilton's principal function,
while there are also several types of corrections due to polymerization.
We conclude that the polymerized improved
action $\Gamma_{\textrm{poly}}$ with the above counter-term
makes it possible to proceed
with the saddle point approximation if desired, however due to presence of  additional
terms proportional to the powers of 
$\lambda$ and the modification of the bulk action due to polymerization, it is reasonable to expect that the thermodynamical properties of the system will be modified in case of
e.g. a black hole. 

\subsection{The model with discrete $p$ \label{sec:eff-poly-without-HJ}}

In this polarization, the kinetic term can simply represented as  $\frac{\hat{p}^{2}}{2m}$.
The problem here is how to represent the potential. It is not clear
how Thiemann's regularization can be used in this case. The reason
is that generally this regularization is used to represent a variable that is discrete and not bounded. To see this, we note that for the regularization to be applied, one should find functions $F(U_{\mu})$ and $G(p)$ such that classically
$\frac{1}{q^{n}}=\left\{ F\left(U_{\mu}\right),G(p)\right\} ^{m}$, with $n,m>0$
so that $F(U_{\mu})$ and $G(p)$  admit a simple representation on Hilbert space. Finding such classical functions may not be very hard, for example we have 
$ \frac{2}{\sqrt{\mu}}\left\{ \sqrt{\big|-i\ln\left(U_{\mu}\right)\big|} ,p\right\}  =\frac{1}{\sqrt{|q|}}$.
However the ability to represent such functions on the Hilbert space is the hard part. Nevertheless, we propose a  scheme for the effective potential based on the observation
that $q^{2}$ in this polarization is replaced by the operator $\widehat{q_{\mu}^{2}}=\frac{\hbar^2}{\mu^2}\left(2-\hat{U}_{\mu}-\hat{U}_{-\mu}\right)$ whose action on a $|q\rangle$ basis is $\widehat{q_{\mu}^{2}}|q\rangle= \frac{4\hbar^2}{\mu^{2}}\sin^{2}\left(\frac{\mu q}{2\hbar}\right)|q\rangle.
$
Based on this, our proposed replacement of the inverse square 
operator is $\widehat{\frac{1}{q_{\mu}^{2}}}|q\rangle=\frac{\mu^{2}}{4\hbar^{2}}\csc^{2}\left(\frac{\mu q}{2\hbar}\right)|q\rangle$
and using this and the  path integral methods,
the effective potential becomes 
\begin{equation}
W^{h}=\frac{\mu^{2}k}{4\hbar^{2}}\csc^{2}\left(\frac{\mu q}{2\hbar}\right).\label{eq:heur-potential}
\end{equation}
This also leads to modifications to the thermodynamics
of the system due to the modification of the bulk action by polymerization as well as absence of any counter-term. This is particularly interesting if the system under study is a black hole.

\section{Discussion\label{sec:conclude}}

In this work we have studied the issue of ill-defined saddle point  approximation that occurs for some systems including
dilatonic black holes and some mathematically analog  mechanical models. This problem
arises due to the fact that the action is not functionally differentiable or diverges on classical
solutions. This issue is rather important since one of the
main methods of deriving thermodynamical properties of such models
is through the saddle point approximation to the path integral that
also can be interpreted as the partition function of the system. The
common solution to this problem is to add a boundary counter-term
to the action, which is a solution to the corresponding Hamilton-Jacobi
equation. A very interesting observation is that with this term
one gets the correct thermodynamics for certain black hole systems. 

Here we seek an alternative method to attack this issue through polymer quantization and by analyzing an analog model exhibiting similar technical issues. The effects of polymerization
may lead to two outcomes: either we will not need to
add any boundary term to the action due to polymer effects,
or we still need to add such a term, but it
is modified by polymer effects. In both cases, it is likely that the process
of polymer quantization will change the thermodynamics.

We first show how this system is modified under different polarizations
of polymer quantization. It turns out that in the polarization where
$q$ is discrete, we still need to add  the boundary term to the action, but the advantage is that the potential can be rather
easily represented on a Hilbert space using the Thiemann's regularization
\cite{Thiemann1998}. This is the case in which the Hamilton-Jacobi
boundary term is modified by polymerization. We then proceed to compute
the effective action and thus derive the effective Hamiltonian using
path integral formulation and then derive the associated  boundary term and show that the classical terms of the polymerized
case match exactly the non-polymerized case while there are corrections
to this counter-term that come from the polymer quantization. This
supports our claim that polymer quantization will change the thermodynamics
of the system due to the polymer modifications of the bulk action
and the  boundary term.

On the other hand, in the polarization where $p$ is discrete, we
argue that in general there should be no need to add a counter-term to the action since
the variable $q$ is bounded due to polymer effects and thus the action
remains functionally differentiable and finite in accordance with
all the allowed variations. However, since the representation of the
potential in this case cannot be pursued directly,
we propose an effective form that replaces the classical potential $W=\nicefrac{1}{q^{2}}$
based on the form of the polymer operator $\widehat{q^{2}_{\mu}}$. Using this, we show that the issues are resolved and the effects of polymer
quantization, render the system well-defined
for saddle point approximation. 


\section*{Acknowledgments}
The authors would like to thank D. Grumiller for his enlightening
comments on \cite{Grumiller2007a}. They would also like to acknowledge the partial support
of CONACyT grant  237351: Implicaciones F\'{i}sicas de la Estructura
del Espaciotiempo. S.R. would like to acknowledge the support of the PROMEP postdoctoral fellowship (through UAM-I) and the grant from
SNI of CONACyT. D.H.O.B. would like
to acknowledge the support of CONACyT grant 283451.

\bibliographystyle{ws-procs975x65}
\bibliography{Poly-bib}

\end{document}